\documentclass{article}

\usepackage[final]{neurips_2021}
\usepackage{graphicx}
\usepackage{natbib}

\usepackage[utf8]{inputenc} 
\usepackage[T1]{fontenc}    
\usepackage{hyperref}       
\usepackage{url}            
\usepackage{booktabs}       
\usepackage{amsfonts}       
\usepackage{nicefrac}       
\usepackage{microtype}      
\usepackage{xcolor}         

\title{Forecasting Global Weather \\ with Graph Neural Networks 
}

\author{%
  Ryan Keisler\\
  \texttt{rkeisler@gmail.com} 
}

\begin{document}

\newcommand*{\era}{\rm{ERA5}}
\newcommand*{\ecmwf}{\rm{ECMWF}}
\newcommand*{\gfs}{\rm{GFS}}

\newcommand*{\jax}{\texttt{jax}}
\newcommand*{\zarr}{\texttt{Zarr}}

\maketitle

\begin{abstract}
  We present a data-driven approach for forecasting global weather using graph neural networks.  The system learns to step forward the current 3D atmospheric state by six hours, and multiple steps are chained together to produce skillful forecasts going out several days into the future.  The underlying model is trained on reanalysis data from ERA5 or forecast data from GFS.  Test performance on metrics such as $Z500$ (geopotential height) and $T850$ (temperature) improves upon previous data-driven approaches and is comparable to operational, full-resolution, physical models from GFS and ECMWF, at least when evaluated on 1-degree scales and when using reanalysis initial conditions.  We also show results from connecting this data-driven model to live, operational forecasts from GFS.
\end{abstract}

\section{Introduction}
\label{sec:introduction}

Numerical weather prediction (NWP), as part of the broader weather enterprise, has had an enormous and positive impact on society.  Decades of steady improvements in the quantity and types of observational data, better modeling techniques, and more computational power have resulted in increasingly accurate weather forecasts and growing adoption of NWP in real-world applications.

While statistical techniques have been used within NWP for decades, the core dynamical engines of these models continue to be based on the physical principles governing the atmosphere and ocean.  More recently, spurred on by advancements in machine learning (ML), there has been a surge of interest in statistical, data-driven techniques for weather forecasting.  The motivation for using ML is to improve upon an already extremely successful NWP program through some combination of better forecasts, faster forecasts, or more forecasts, i.e. larger ensembles.  There may also be opportunities for using ML to advance our scientific understanding of the underlying physical processes \citep{cranmer2020_discovering_symbolic}.

There is currently a very active hub of research at the intersection of NWP and ML.  Example research areas include: faster or more accurate solving of relevant PDEs \citep{bar-sinai2019_learning_discretizations_hoyer, alieva2021_ml_fluid_dynamics_hoyer, um2021solverintheloop_thuerey, li2021_fourier, brandstetter2022_message_pdes}, data assimilation \citep{frerix2021_variational_data_assimilation, maulik2021_variational_data_assimilation}, subgrid parameterization \citep{brenowitz2019_parametrization, chantry2021_gravity_parameterization, yuval2021_parameterization, meyer2021_3dcloud_dueben_parameterization}, nowcasting \citep{sonderby2020_metnet1, agrawal2019_google_nowcasting, espeholt2021_google_nowcasting_metnet2, klocek2021_msnowcasting, ravuri2021_deepmind_nowcasting}, nudging global NWP models \citep{watt-meyer2021_nudging}, replacing global NWP models \citep{dueben2018_global, arcomano2020_reservoir, rasp2021_resnet, scher2021_spherical, weyn2020_cnn_cubed_sphere, weyn2021_subseasonal_ensemble, clare2021_distribution}, high-level proposals for end-to-end ML weather systems \citep{schultz2021_can_deep_learning_beat_nwp}, and decision making in the context of high-impact weather events \citep{mcgovern2017_ai_high_impact_weather, gagne2017_hail}.

In this work we present a data-driven, machine learning model for forecasting global weather.  Our approach is similar to and inspired by previous efforts to use ML to emulate global NWP such as \cite{rasp2021_resnet} and \cite{weyn2020_cnn_cubed_sphere}, but with some notable differences described below.

First, we model a significantly denser physical system.  Each step of our model outputs 6 physical variables defined on 13 pressure levels on a 1.0-degree latitude/longitude grid, corresponding to $\sim$5,000,000 physical quantities.  For reference, this is $\sim$50-2000X larger than the number of physical quantities predicted by the models in \cite{rasp2021_resnet} or \cite{weyn2020_cnn_cubed_sphere}.  The motivation for modeling a denser physical system was to reach closer towards the highly successful and much denser physical system being modeled in traditional NWP.  Put another way, our system uses ML to perform non-linear interpolation between previously seen spatiotemporal patterns, and if the extent of that pattern in space and time is small, then it will have simpler dynamics that are easier to interpolate between.  In the limit of using the extremely dense spatial grid and short time step used in operational NWP simulations (well beyond the scope of this work), the ML system would "only" have to learn the relatively compact set of physical laws driving those systems.

Second, our approach differs from \cite{rasp2021_resnet} and \cite{weyn2020_cnn_cubed_sphere} in that we use message-passing graph neural networks (GNNs) rather than convolutional neural networks (CNNs).  The strongest motivation for this choice was observing the success that \cite{pfaff2021_meshgraphnets} had using GNNs for simulating physical systems.  Message-passing GNNs provide a more general architecture than conventional CNNs, and we believe the flexibility of the underlying spatial graph provides benefits that are well suited to NWP.  These include (i) handling the spherical geometry of the earth in a straightforward way, (ii) the potential for learning multi-resolution models, which we explored briefly in this work, and (iii) the potential for adaptive mesh refinement, i.e. putting the compute where it is needed at each time step, which was explored in \cite{pfaff2021_meshgraphnets} but not in this work.

The paper is structured as follows.  In Section~\ref{sec:data} we describe the data used to train and validate our model.  In Section~\ref{sec:model} we describe the architecture and training of our model.  We cover results in Section~\ref{sec:results}, provide additional disccusion in Section~\ref{sec:discussion}, and conclude in Section~\ref{sec:conclusion}.

\section{Data}
\label{sec:data}
We train and validate our model using one of two datasets: the \era~reanalysis dataset from the European Centre for Medium-Range Weather Forecasts (\ecmwf) and a subset of 2021 forecasts from NOAA's Global Forecast System (\gfs).

\subsection{ERA5}
\label{sec:era_data}
For most of the results show in this work we use the \era~reanalysis dataset \citep{hersbach2020_era5} from \ecmwf.  This dataset synthesizes decades of meteorological observations into a consistent physical framework using 4D-Var data assimilation \citep{rabier2000_4dvar}.  The result is a set of hourly snapshots of the 3D atmospheric state from 1950 to present.

We use six physical variables: temperature $T$, geopotential height $Z$, specific humidity $Q$, the eastward component of the wind $U$, the northward component of the wind $V$, and the vertical component of the wind $W$.  These variables are interpolated onto a 3D grid prior to download: a 1-degree latitude/longitude grid (horizontal resolution) and 13 pressure levels (vertical resolution).  The pressure levels are 50, 100, 150, 200, 250, 300, 400, 500, 600, 700, 850, 925, and 1000 hPa.  We download data for every third hour from 1979-01-01 to 2021-01-01.  After downloading and processing data from the Copernicus Climate Data Store, we persist the data as a single \zarr~array, backed by Google Cloud Storage, with dimensions $(n_{\rm{time}}, n_{\rm{lat}}, n_{\rm{lon}}, n_{\rm{variables}}*n_{\rm{levels}}) = (122728, 181, 360, 78)$.  We divide the data into three sets: validation (1991, 2004, 2017), testing (2012, 2016, 2020), and training (all other years in 1979-2020, inclusive).

\subsection{GFS}
\label{sec:gfs_data}

As discussed more in Section~\ref{sec:results_gfs}, we also explore training  on historical forecasts from NOAA's \gfs.  We use 712 forecasts issued by GFS v16 between 2021-04-23 and 2021-10-18.  We keep every third hour of forecast output (f000, f003, etc.) from the first 249 hours of output, for a total of 59,096 individual time steps.  We divide the data into a set of 18-contiguous-day chunks for training and a set of 6-day-contiguous chunks for validation, with 3-day buffers between all chunks.

We use the same set of physical variables, the same 1-degree latitude/longitude grid, and the same 13 pressure levels that we use with \era.  We note that, despite these similarities between our \era~and \gfs~datasets, we do observe subtle differences between them (e.g. how variables are interpolated "underground" for pressure levels that are larger than the surface pressure) that prevent straightforward interoperability between \era~and \gfs.

\subsection{Additional Datasets}
We use three additional datasets as inputs to the model.  First, we use two static datasets from \era: a land-sea mask and an orography dataset, both at 1-degree resolution.  Additionally, we use an analytic approximation to the top-of-atmosphere solar radiation at each location, evaluated at the current time and 10 neighboring times spanning $\pm$ 12 hours.

\section{Model}
\label{sec:model}

Our data-driven weather forecasting model is based on message-passing graph neural networks (GNN) and draws heavily from the approach of \cite{pfaff2021_meshgraphnets}.  The parameters of the GNNs are machine-learned using historical training data.  We use \jax~and two \jax-based packages: \texttt{jraph} for building the graph neural network structures and \texttt{haiku} for managing the actual neural networks.

\subsection{Architecture}

\begin{figure}
  \centering
  \includegraphics[width=1.0\columnwidth]{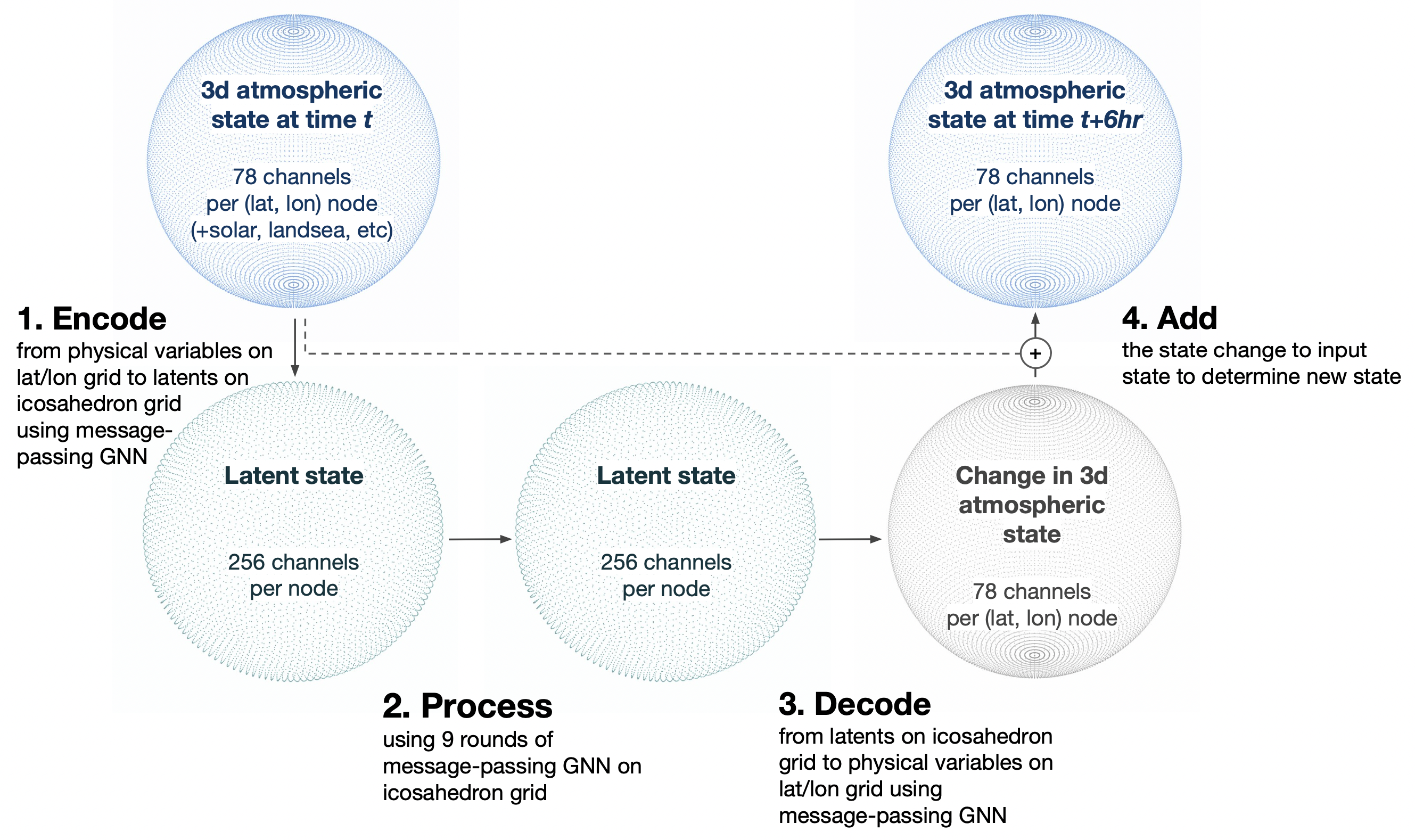}
  \caption{Using the current atmospheric state, the model evolves the state forward by 6 hours.  The 3D atmospheric state is defined on a uniform latitude/longitude grid, with 78 channels per pixel (6 physical variables $\times$ 13 pressure levels = 78 channels).  An Encoder GNN encodes onto latent features defined on a icosahedron grid, a Processor GNN performs additional processing of the latents, and a Decoder GNN maps back to the atmospheric state on a latitude/longitude grid.}
  \label{fig:arch}
\end{figure}

The model architecture, shown in Figure~\ref{fig:arch}, is made up of three components: an Encoder, a Processor, and a Decoder.  At a high level, the Encoder maps from the native data space (physical data on a latitude/longitude grid) to an intermediate space (abstract feature data on an icosahedron grid).  The Processor then processes in that intermediate space, and the Decoder maps back to the native data space.

The motivation to use an icosahedron grid for the intermediate representation is that it would provide a processing grid that is more uniformly distributed and efficient than the original latitude/longitude grid.  We use the \texttt{h3} package\footnote{\url{https://h3geo.org/}} to define a \texttt{level=2} icosahedron grid that has 5,882 nodes (compared to 65,160 in the original lat/lon grid) with $\sim$3-degree ($\sim$330 km) angular separation between nodes.

\begin{figure}
  \centering
  \includegraphics[width=1.0\columnwidth]{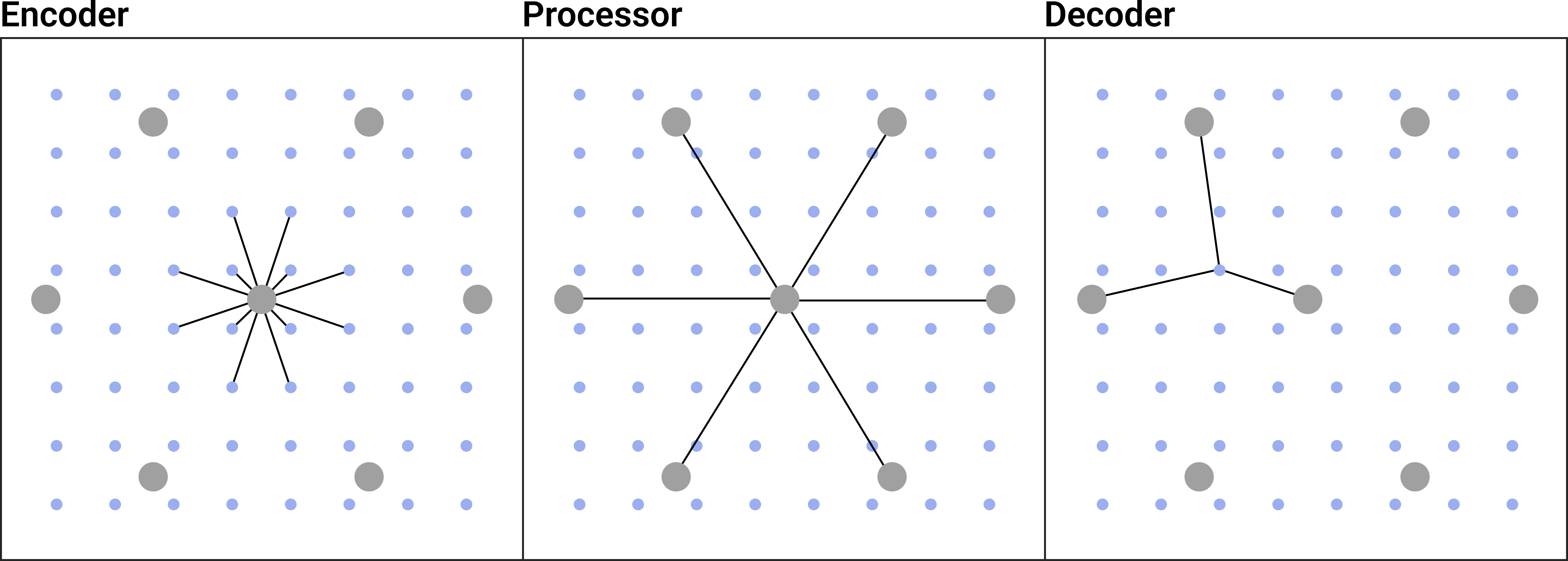}
  \caption{A schematic view of the local graph connectivity in the Encoder, Processor, and Decoder.  Left: local spatial and channel information is encoded into an icosahedron node using data from nearby nodes on the input latitude/longitude grid.  Center: data on the icosahedron node is further processed using data from nearby icosahedron nodes (including itself, which is not explicitly shown).  Right: the output latitude/longitude data is created by decoding data from nearby icosahedron nodes.}
  \label{fig:graph}
\end{figure}

Each of the three model components is implemented as a message-passing GNN.  We refer the reader to \citet{pfaff2021_meshgraphnets} or \cite{battaglia2018_gnn_overview} for more information, but in short, these GNNs are characterized by an underlying graph (i.e. a set of nodes plus a set of directional edges connecting some of those nodes), a neural network that updates node features, and a neural network that updates edge features.  For our neural networks we use 2-layer MLPs with ReLU activation, LayerNorm, and 256 output channels.  In Figure~\ref{fig:graph} we provide a schematic view of the local graph connectivity in the Encoder, Processor, and Decoder.

\subsubsection{Encoder} As mentioned above, the Encoder maps from physical data defined on a latitude/longitude grid to abstract latent features defined on an icosahedron grid.  The Encoder GNN uses a bipartite graph (lat/lon$\rightarrow$icosahedron) with edges only between nodes in the lat/lon grid and nodes in the icosahedron grid.  Put another way, spatial and channel information in the local neighborhood of each icosahedron node is gathered using connections to nearby lat/lon nodes.

The initial node features are the 78 atmospheric variables described in Section~\ref{sec:era_data}, plus solar radiation, orography, land-sea mask, the day-of-year, $\sin({lat})$, $\cos({lat})$, $\sin({lon})$, and $\cos({lon})$.  The initial edge features are the positions of the lat/lon nodes connected to each icosahedron node.  These positions are provided in a local coordinate system that is defined relative to each icosahedron node.

\subsubsection{Processor} The Processor iteratively processes the 256-channel latent feature data on the icosahedron grid using 9 rounds of message-passing GNNs.  During each round, a node exchanges information with itself and its immediate neighbors.  There are residual connections between each round of processing.

\subsubsection{Decoder} The Decoder maps back to physical data defined on a latitude/longitude grid.  The underlying graph is again bipartite, this time mapping icosahedron$\rightarrow$lat/lon.  The inputs to the Decoder come from the Processor, plus a skip connection back to the original state of the 78 atmospheric variables on the latitude/longitude grid.  The output of the Decoder is the predicted 6-hour change in the 78 atmospheric variables, which is then added to the initial state to produce the new state.  We found 6 hours to be a good balance between shorter time steps (simpler dynamics to model but more iterations required during rollout) and longer time steps (fewer iterations required during rollout but modeling more complex dynamics).

\subsection{Model Size and Latency}
The model has 6.7M parameters, i.e. 27 MB of information when using float32 representation.  This is $\sim$80,000X smaller than the 2.1 TB of \era~data used to train the model.  As expected, we see no evidence of overfitting on the training dataset; validation and test losses are indistinguishable from the training loss.

The model is fast to run.  After the initial overhead of loading the model weights and compiling to XLA (all handled by \jax), a single 6-hour model step takes 0.04 seconds when running on a NVIDIA A100 GPU, i.e. creating a 5-day forecast takes 0.8 seconds.

\subsection{Training}
\label{sec:training}
We trained our final model using the Adam optimizer and a 3-round training schedule with progressively smaller learning rates: 3.5 days of training at lr=3e-4, 1 day at lr=3e-5, and 1 day at lr=3e-6.  The total training time was 5.5 days on a single NVIDIA A100 GPU, which cost approximately \$370 using the Google Cloud Platform.  The training procedure used multi-resolution training data, a multi-step loss, and a specific loss normalization, as described below.

 \subsubsection{Multi-resolution training data}
 One useful feature of using message-passing GNNs is that we can encode the relative positions between nodes into the messages, so that a single model can learn from data at different resolutions.  We took advantage of this by first training on 2-degree data for the first round of training and then switching to training on 1-degree data for the last two rounds.  For reasons we do not understand, this produced better results than training on 1-degree data throughout.
 
\subsubsection{Multi-step loss}
Although our core model makes a single, 6-hour step into the future, we would like it to work well when making, say, a 5-day forecast requiring a 20-step rollout.  We encouraged this behavior by training with a multi-step loss.  More specifically, during training we rolled out the model for $\sim$10 steps and accumulated the loss at each of those 6-hour steps.  We used a progressively larger rollout for each round of training: 4, 8, and 12-step losses, corresponding to 1, 2, and 3-day rollouts, for the three rounds of training.  Using even larger rollouts is enticing, but there are probably diminishing returns \citep{metz2021_gradients_are_not_all_you_need}, and in practice we obtained only slightly worse results when using a 4-step loss throughout.

\subsubsection{Loss normalization}
\label{sec:loss}
We observed that one of the most important choices for getting good performance was how we normalized the data when calculating the training loss.  Prior to calculating our mean-squared error (MSE) loss, we re-scale each physical variable such that it has unit-variance in its 3-hour temporal difference.  For example, we divide the temperature data at all pressure levels by $\sigma_{T,3hr}$, where $\sigma_{T,3hr}^2$ is the variance of the 3-hour change in temperature, averaged across space (lat/lon + pressure levels) and time ($\sim$100 random temporal frames).  The motivations for this choice are (i) we are interested in predicting the dynamics of the system, so normalizing by the magnitude of the dynamics is appropriate and (ii) a physically-meaningful unit of error, e.g. 1 K of temperature error, should count the same whether it is happening at the lower or upper levels of the atmosphere.

We also re-scale the data by a nominal, static air density at each pressure level.  This did not have a strong impact on performance, but we did use this re-scaling in our final model.

Finally, when summing the loss across the latitude/longitude grid, we use a weight proportional to each pixel's area, i.e. a $\cos(lat)$ weighting.

\section{Results}
\label{sec:results}

\subsection{Single-step results}
\begin{figure}
  \centering
  \includegraphics[width=1.0\columnwidth]{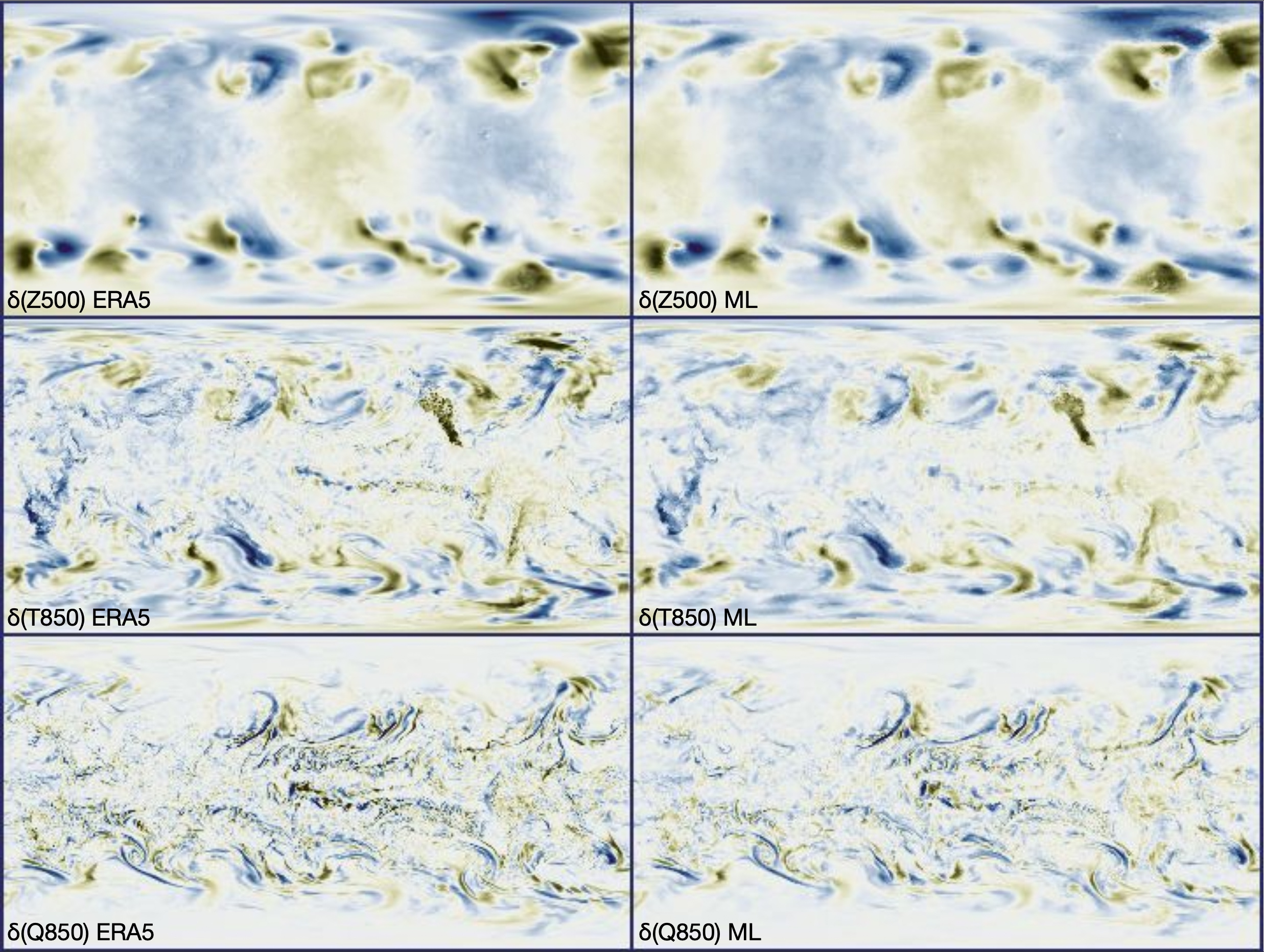}
  \caption{An example of the 6-hour difference in geopotential height, temperature, and humidity in the \era~dataset (left column) and the prediction from the machine learning model (right column).  The model is able to accurately predict 6-hour changes in these variables using only the initial state.  All raster figures in this paper use a $1^\circ$ latitude/longitude grid centered on the prime meridian.}
  \label{fig:6hr_diff_ztq}
\end{figure}

\begin{figure}
  \centering
  \includegraphics[width=1.0\columnwidth]{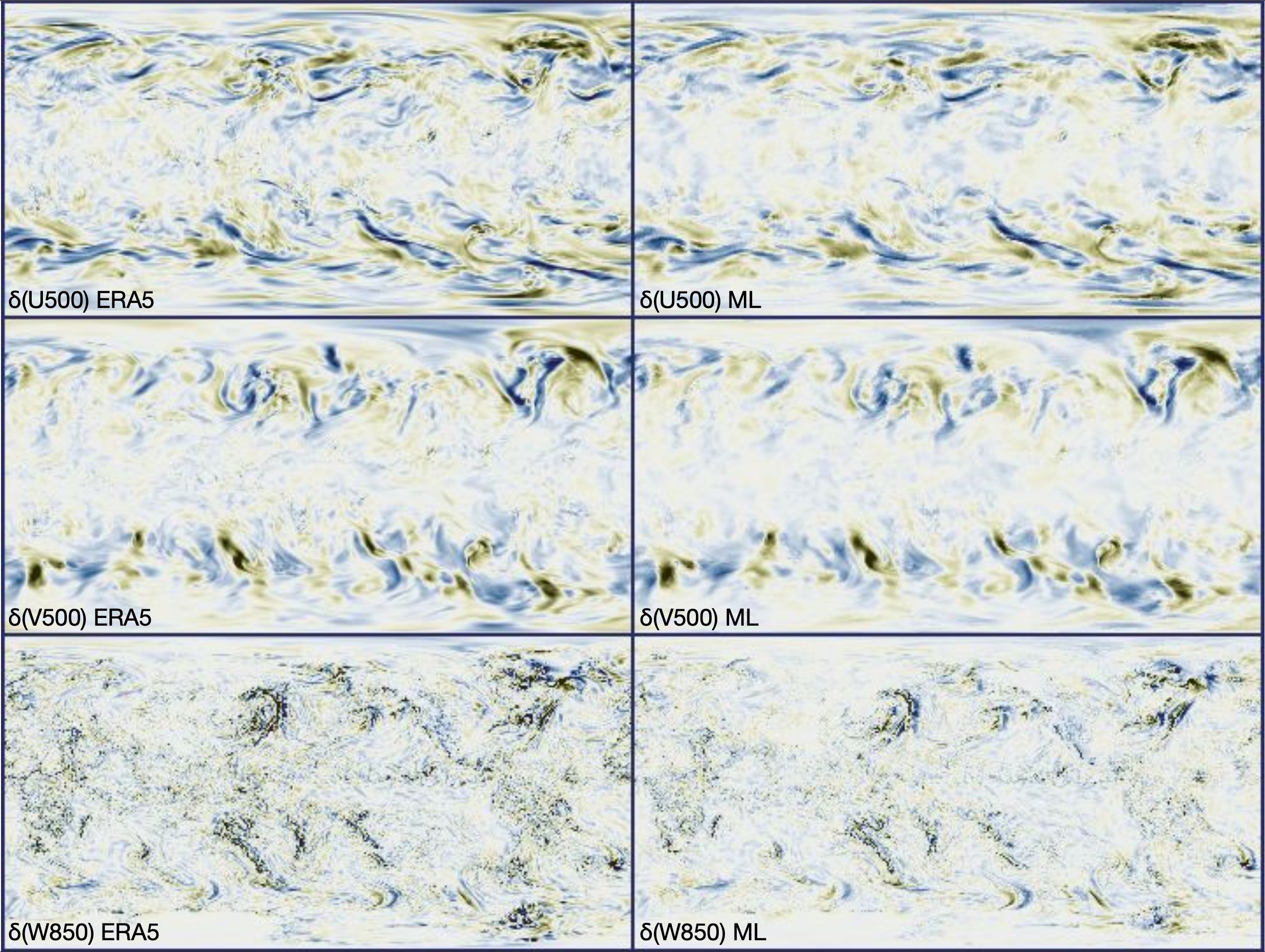}
\caption{An example of the 6-hour difference in eastward component of the wind, the northward component of the wind, and the vertical component of the wind in the \era~dataset (left column) and the prediction from the machine learning model (right column).  The model is able to accurately predict 6-hour changes in these variables using only the initial state.}
\label{fig:6hr_diff_uvw}
\end{figure}

Our model operates by stepping forward in 6-hour steps, and the results from an example step are shown in Figures~\ref{fig:6hr_diff_ztq} and \ref{fig:6hr_diff_uvw}.  We see that the model has learned how to use the current atmospheric state to predict the change in that state over the next 6 hours.

\subsection{Multi-step results}

\begin{figure}
  \centering
  \includegraphics[width=1.0\columnwidth]{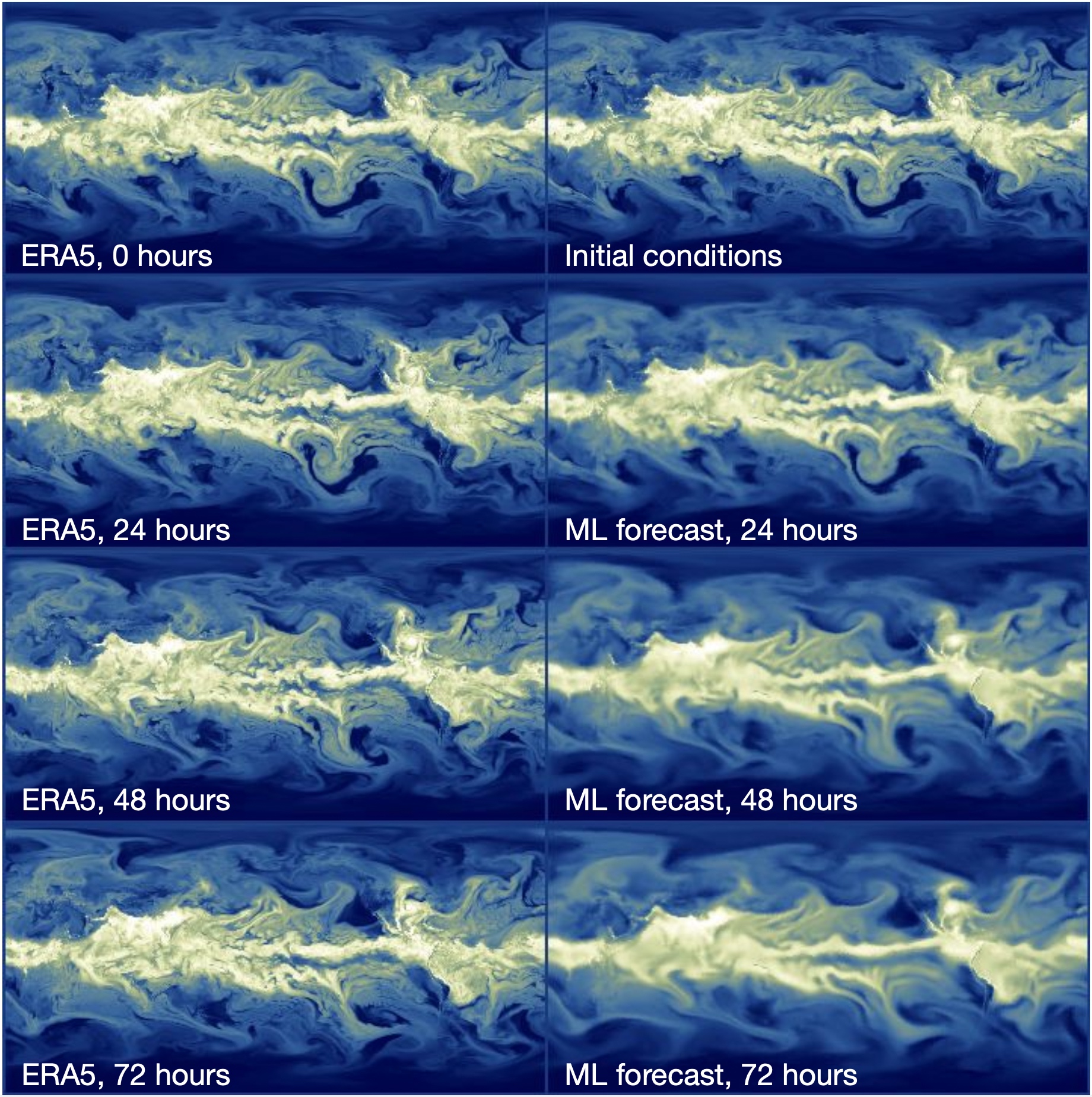}
  \caption{An example multi-step rollout of the ML forecast vs reanalysis data from \era.  Beginning with the \era~initial conditions at 0 hours, the ML system steps forward autoregressively in 6-hour steps.  While the model evolves 78 separate physical channels, we show only Q850, the specific humidity on the 850 hPa pressure level.  The output of the ML forecast generally tracks the large-scale flows seen in \era, although the predictions do become smoother over time.  Additional media, including videos, can be found at \url{https://rkeisler.github.io/graph_weather}}
  \label{fig:rollout}
\end{figure}

We can also look at how the model performs when rolled out for many steps.  The process is autoregressive: the output of the first step of the model becomes the input for the second step of the model, and so on.  In Figure~\ref{fig:rollout} we show the result of a 12-step (3-day) rollout.  We see that the output of the data-driven forecast generally tracks the large-scale flows of the specific humidity at 850 hPa, although the data-driven forecast does become smoother over time.

\subsection{Stability}
We observe that the model forecast is numerically stable when being rolled out $\sim$6 days.  This is somewhat surprising given that we did not enforce any kind of physically motivated "conservation law" (e.g. conservation of momentum) nor did we attempt to stabilize model rollouts by training with additional noise as in \cite{pfaff2021_meshgraphnets}.

However, we did observe hexagon-patterned instabilities beginning to form when rolling out past $\sim$6 days.  This seems to be caused by the use of the icosahedron grid in the Processor; we did not observe this pattern when using a simpler (but slightly less performant) architecture that did not use the icosahedron grid and instead remained in latitude/longitude space.  We speculate that this hexagon instability pattern would be mitigated by some combination of (i) using and predicting multiple time steps as in \cite{weyn2020_cnn_cubed_sphere}, rather than using and predicting only one time step as in this work, (ii) applying additive noise to the inputs of the model as in \cite{pfaff2021_meshgraphnets}, or (iii) mildly randomizing the graph connectivity within the icosahedron grid, a kind of edge Dropout.

\subsection{Comparison to data-driven models}
\label{sec:results_vs_ml}
We compare the forecast performance of our data-driven model to two previous data-driven models: \citet{rasp2021_resnet} and \citet{weyn2020_cnn_cubed_sphere}.  To the best of our knowledge, these are the two best-performing data-driven global weather forecast systems to date.  The methods and discussions presented in these two papers were a major source of inspiration for the work presented here.  

For reference, the best-performing models in \citet{rasp2021_resnet} used a CNN to directly (i.e. not autoregressively) predict a single physical variable, like temperature at 850 hPa, at 5.625-degree resolution.  \citet{weyn2020_cnn_cubed_sphere} used a CNN to autoregressively predict four variables at $\sim$1.9-degree resolution: geopotential height at 500 hPa and at 1000 hPa, the 300-to700-hPa geopotential thickness, and the 2-meter temperature.  The major differences between these models and ours are: (i) we use data with finer spatial resolution (1.0 deg vs 1.9 or 5.625 deg), (ii) we predict 78 channels of information rather than <5 in these works, and (iii) we use GNNs rather than CNNs.  See Section~\ref{sec:introduction} and Section~\ref{sec:discussion} for additional discussions of these differences.

\begin{figure}
  \centering
  \includegraphics[width=1.0\columnwidth]{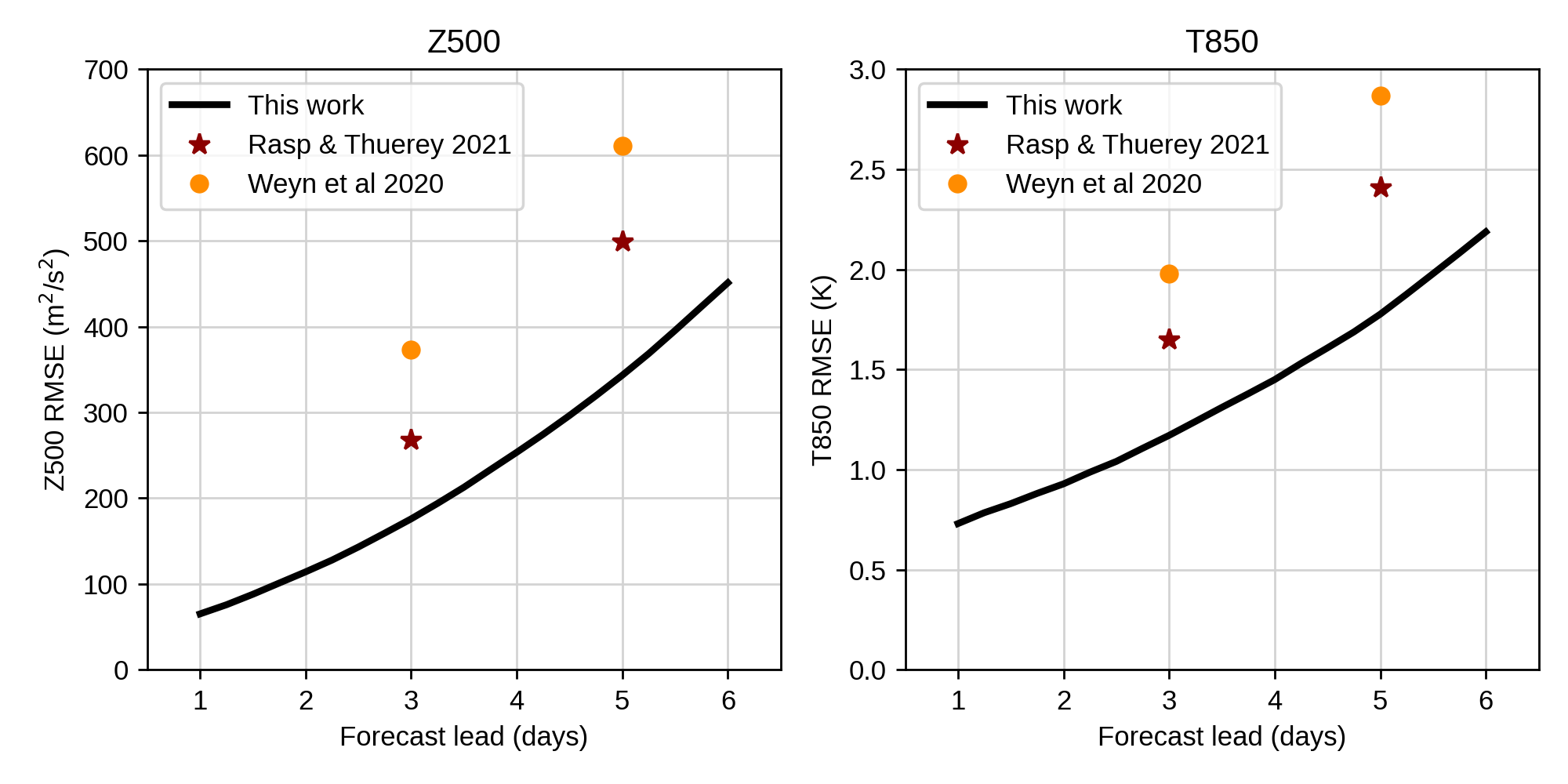}
  \caption{Performance metrics (RMSE of geopotential height and temperature) of our data-driven model vs previous data-driven models from \citet{rasp2021_resnet} and \citet{weyn2020_cnn_cubed_sphere}.  The metrics for our model are calculated on a 1.0$^\circ$ latitude/longitude grid with FWHM=1.5$^\circ$ spherical gaussian smoothing.  All metrics are computed globally (i.e. extratropics and tropics) over the year 2016.}
  \label{fig:metrics_ml}
\end{figure}

We take performance metrics for these two models from the WeatherBench \citep{rasp2020_weatherbench} website\footnote{\url{https://github.com/pangeo-data/WeatherBench}}.  We attempt to replicate the benchmarking process used in these works: we evaluate our model performance on 2016 data, and we evaluate globally (i.e. tropics + extratropics).  For reasons described in the following section, we calculate metrics for our data-driven model on a 1.0-degree latitude/longitude grid with FWHM=1.5-degree spherical gaussian smoothing.  Results are shown in Figure~\ref{fig:metrics_ml}.  We see that the model presented in this work outperforms these two previous approaches, in addition to (and perhaps due to) modeling more channels of information at higher spatial and temporal resolution.

\subsection{Comparison to operational NWP models}
\label{sec:results_vs_operational}
We compare the forecast performance of our data-driven model to two operational, physics-based NWP models: ECMWF's IFS and NOAA's GFS.  Below we describe some limitations of this comparison before discussing the actual results of the comparison.

We use the 2020 performance metrics for ECMWF and GFS provided by WMO-LCDNV\footnote{\url{https://apps.ecmwf.int/wmolcdnv/}}.  These metrics represent the operational, full-resolution models evaluated on a 1.5-degree latitude/longitude grid with T120 spectral truncation.  We calculate metrics for our data-driven model on a 1.0-degree latitude/longitude grid with FWHM=1.5-degree spherical gaussian smoothing to approximate the T120 truncation.  We compute metrics for our model only over the extratropics in order to compare to the average of the northern and southern hemisphere metrics provided by WMO-LCDNV.

It should be said that while evaluating performance metrics on these relatively coarse angular scales is standard practice, ECMWF and GFS run simulations and provide data products at significantly finer spatial resolution, e.g. simulations at $\sim$10 km horizontal resolution and data products at 0.25 degrees.

\begin{figure}
  \centering
  \includegraphics[width=1.0\columnwidth]{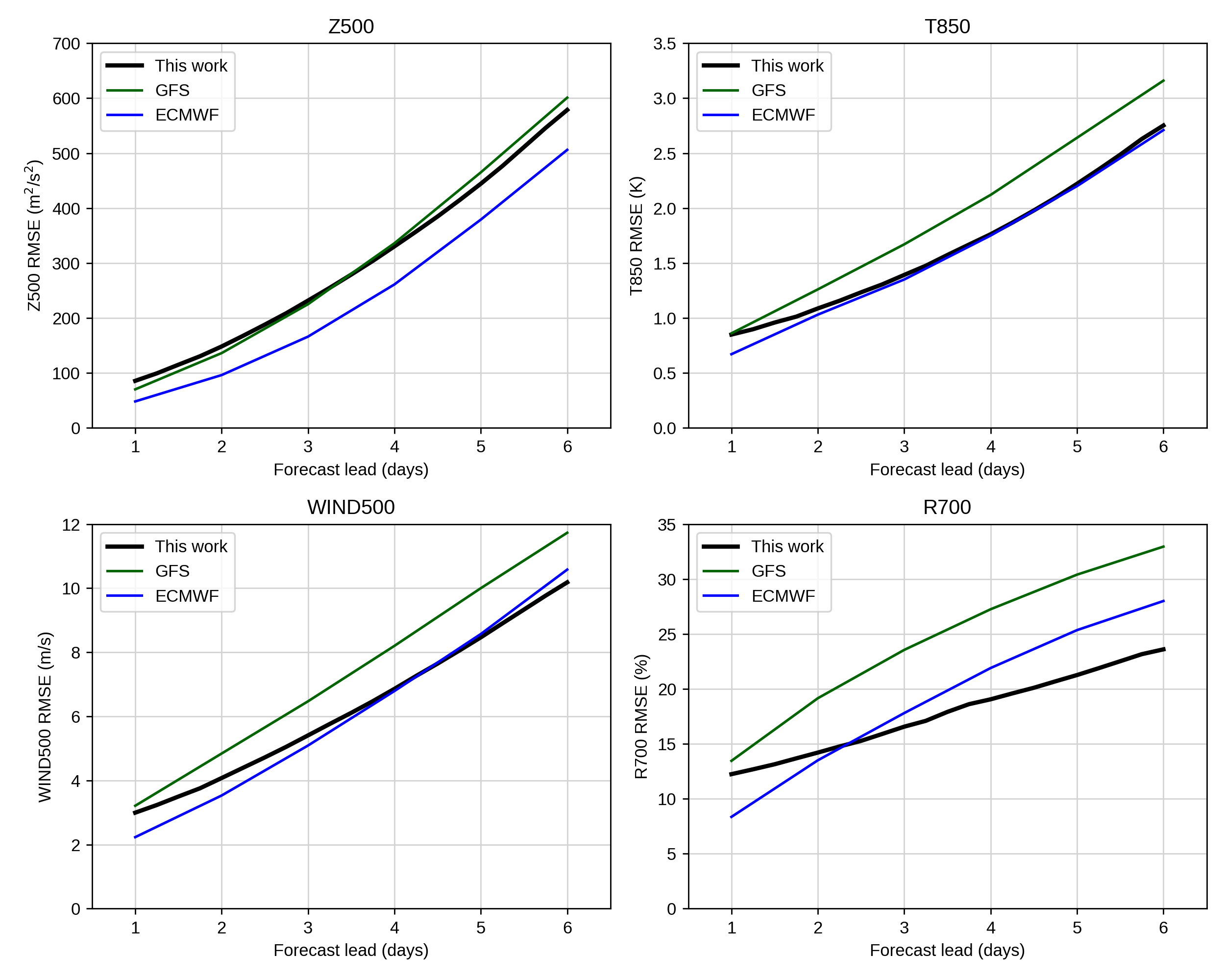}
  \caption{Performance metrics (RMSE of geopotential height, temperature, wind, and relative humidity) of our data-driven model vs operational, physics-based models.  The data-driven model generally outperforms the GFS v15.2 model used in production in 2020 and generally underperforms the ECMWF model used in production in 2020.  The metrics for GFS and ECMWF are taken from \href{https://apps.ecmwf.int/wmolcdnv/}{WMO-LCDNV} and represent the operational, full-resolution models evaluated on a 1.5$^\circ$ latitude/longitude grid with T120 spectral truncation.  The metrics for our data-driven model are calculated on a 1.0$^\circ$ latitude/longitude grid with FWHM=1.5$^\circ$ spherical gaussian smoothing to approximate the T120 truncation.  Additionally, the metrics for our data-driven model have been shifted to the left by 0.25 days to approximately remove the artificial advantage gained by the 0.5-day data assimilation window of the \era~data used to initialize those forecasts.  All metrics are computed over the year 2020 and only over the extratropics.}
  \label{fig:metrics_phys}
\end{figure}

It should also be said that the performance metrics for ECMWF and GFS are evaluated as true forecasts, while we are making our data-driven forecasts after the fact.  Additionally, the initial conditions for our data-driven model are from \era, a reanalysis dataset that assimilates data in 12-hour windows.  This means that the initialization of a particular data-driven forecast could contain information from 0 to 12 hours into the future.  To approximately correct for this artificial advantage, we have shifted the data-driven metrics to the left by 6 hours in Figure~\ref{fig:metrics_phys}

With these caveats, we can see in Figure~\ref{fig:metrics_phys} that the data-driven model performs remarkably well.  It generally performs better than the GFS v15.2 model used in production in 2020 and is comparable in performance to the ECMWF model used in production in 2020.

\subsection{Connecting to live GFS runs}
\label{sec:results_gfs}
The work presented up to this point has focused on building and validating a data-driven weather forecast model using historical reanalysis data from \era.  In this section we demonstrate that this kind of model can also be connected to a live, operational weather forecast system.  The motivations for doing this are (i) to simply demonstrate that our data-driven model can, without much additional effort, be fused with a live system, producing a live, hybrid physics+ML system, and (ii) to demonstrate that this system can anticipate forecast-to-forecast changes (e.g. changes between a 06Z forecast and a 12Z forecast) prior to those changes being published, which could be useful for time-sensitive applications.

We target the operational Global Forecast System (GFS) model from NOAA, because its forecast data is publicly and freely available.  We train a model on historical GFS v16 forecasts made in April-October 2021.  Our GFS-trained, data-driven model is essentially an emulator for the physics-based GFS forecast engine.  It takes as input any single GFS data snapshot, e.g. the 0-hour (f000) or the 120-hour (f120) forecast data product, and quickly (within $\sim$1 second) extends that forecast into the future by some number of days.  We can view the combined system as a hybrid physics+ML model, in which the data assimilation step and the first forecast steps are performed by the physics-based GFS model, and additional forecast steps are performed by the data-driven model.

The hybrid system works as follows.  The GFS model publishes its forecast data products as usual.  We then have a simple Python script scanning for and downloading new data as it is made available.  Once a new forecast snapshot, e.g. f120, is downloaded, we preprocess it and use it to initialize a multi-step rollout of our data-driven forecast model.  The time horizon of this rollout, e.g. 3 days, is a free parameter which can be used to balance an inherent tradeoff between wanting a longer lead time with respect to the published data (i.e. more time to make a decision) and wanting a more accurate approximation to the true forecast-to-forecast change (i.e. making a higher quality decision).  By repeating this process for each new forecast snapshot (f000, f003, etc.) the data-driven model is able to "look ahead" and anticipate where the GFS forecast is heading before it gets there.

\begin{figure}
  \centering
  \includegraphics[width=1.0\columnwidth]{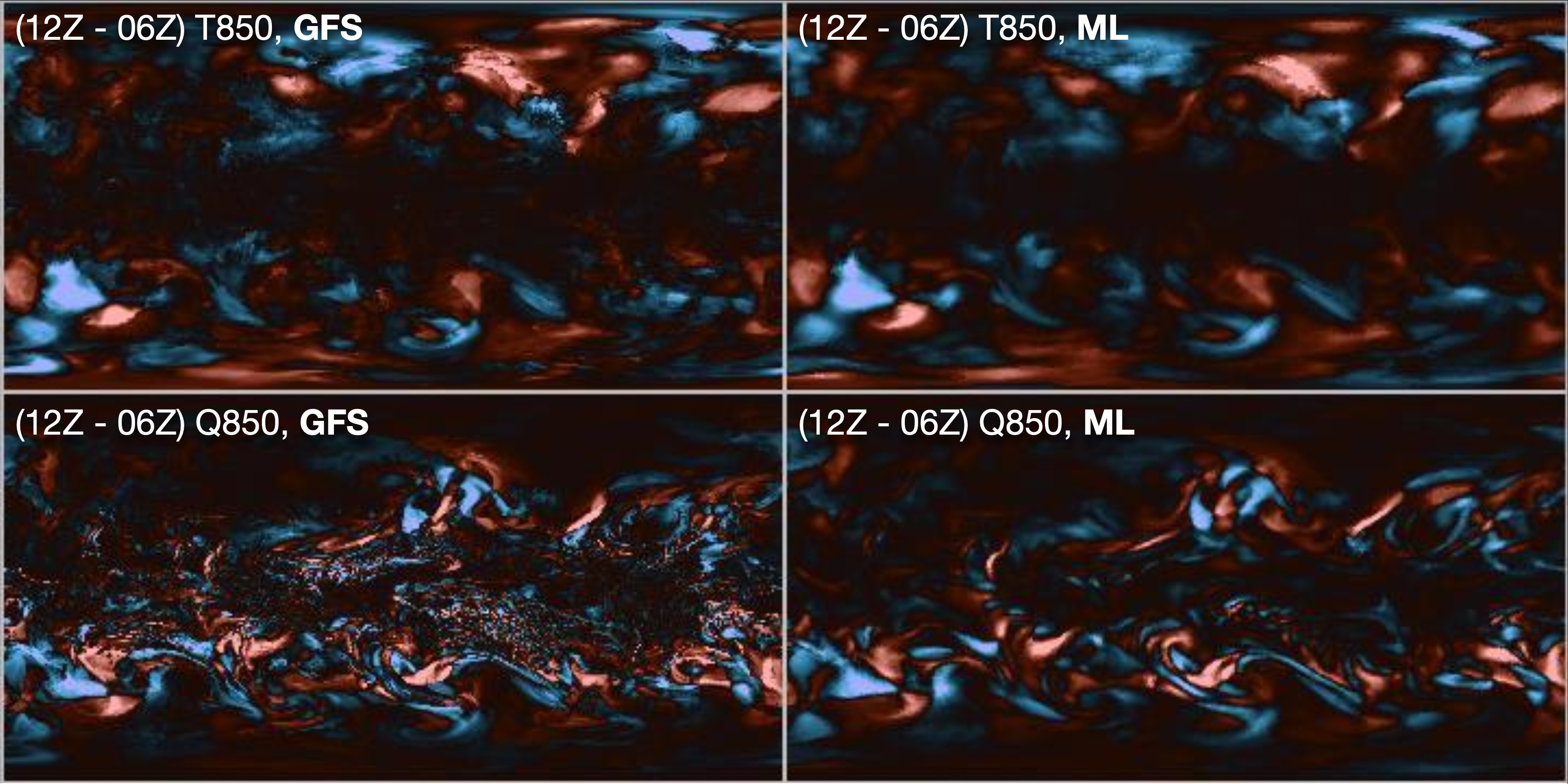}
  \caption{An example of connecting our data-driven model to live data from GFS.  We show forecast-to-forecast changes in temperature (top row) and humidity (bottom row) from two sets of \gfs~forecasts: the 06Z and 12Z forecasts from 2022-01-25, with the forecast valid for 2022-02-04T06:00:00.  The data-driven model (right column) was connected to the live output of these GFS forecasts and accurately approximated the true 06Z-to-12Z change (left column) 22 minutes prior to the true change being published.}
  \label{fig:prod_gfs}
\end{figure}

In Figure~\ref{fig:prod_gfs} we show an example of connecting our data-driven model to live GFS data.  In this example we tracked changes in temperature and humidity made between two forecasts, the 06Z and 12Z forecasts on 2022-01-25.  The data-driven model was configured to extend the forecast by 3 days, corresponding to a lead time of $\sim$20 minutes (because it takes GFS $\sim$20 minutes to publish 3 days worth of forecast data).  We see that this setting gave a high-correlation approximation to the true forecast-to-forecast change, and, in this particular case, did so 22 minutes earlier than the relevant 12Z GFS data was published.  Whether or not this level of accuracy and lead time would be useful would depend on the actual application, but regardless, we find it interesting that the data-driven model is able to anticipate where the physics-based model is going before it actually gets there.

\section{Discussion}
\label{sec:discussion}
In this section we discuss why we believe the data-driven model described in this work performs well,
i.e. why it outperforms previous data-driven approaches and why, with the caveats described in Section~\ref{sec:results_vs_operational}, it is comparable in performance to operational models.  Our work was aimed at optimizing forecast performance within our time and compute budget, and we did so using a large number of small, intuition-building experiments.  As such, we did not prioritize clarity of understanding and cannot point to a set of clean ablation studies.  Nonetheless, we feel it is worthwhile to provide our experiment-informed but somewhat subjective view of why this model performs well.

\begin{itemize}
  \item \textbf{GNNs} - We have highlighted our use of message-passing graph neural networks.  This design choice was motivated by the success of \cite{pfaff2021_meshgraphnets} in simulating physical systems with GNNs.  We have mild empirical evidence from an early experiment that the ability to aggregate information on the sphere over a physically-uniform neighborhood (e.g. 100 km) improves upon using the latitude-dependent neighborhood of, say, a 3x3 CNN kernel.  We have not tested that the generality provided by message-passing GNNs is in itself beneficial, but we believe it would be critical for future work such as adaptive meshing or multi-resolution models; in these applications, a node would need to communicate to its neighbors "here is my data" but also "here is my location".
  
  \item \textbf{Simulating a dense physical system} - Because traditional NWP works extremely well, we were motivated to design a data-driven model in which key prognostic variables (temperature, wind, etc.) are evolved forward with relatively short time steps on a relatively dense 3D grid.  As discussed below, we model a system that is denser than previous data-driven approaches, but we note that our model is still orders of magnitude coarser than operational NWP simulations, roughly 10X in each spatial dimension and 100X in the time step.
  
  Here we discuss in more detail the similarities and differences between our approach and that of \cite{weyn2020_cnn_cubed_sphere} and \cite{rasp2021_resnet}.  Like \cite{weyn2020_cnn_cubed_sphere}, we use an iterative, auto-regressive model with 6-hour time steps, and we train with a multi-step loss, but we model 78 physical channels (6 physical variables on 13 pressure levels) rather than 4 physical channels, and we train a ML model with 10X more parameters.  Like \cite{rasp2021_resnet}, we provide $\sim$100 input channels (key prognostic variables defined on $\sim$10 pressure levels) to a $\sim$5M-parameter model, but we do so with significantly finer horizontal resolution (1.0 vs 5.625 degrees) and we carry these channels forward auto-regressively rather than directly making multi-day forecasts.  The end result is that our model carries forward $\sim$50-2000X more information than these previous data-driven approaches.  We believe that our decision to simulate a denser physical system was likely partially responsible for the improvement in forecast performance that we observe relative to these approaches.
  
  We do acknowledge that the picture is not perfectly clear.  For example, with other design choices (model size, spatial resolution, etc.) held fixed, we found that a 6-hour time step yielded better multi-day forecast performance than a 3-hour time step, presumably because there were twice as many error-accumulating steps during rollout in the 3-hour case.\footnote{On the other hand, a 6-hour time step also outperformed a 12-hour time step, presumably because the dynamics are simpler to model over the shorter 6-hour time scale.  In the end we chose to use a 6-hour time step.}  And as described in Section~\ref{sec:training}, for reasons we do not understand, we found that an initial round of training on coarser, 2-degree data improved the performance of our 1-degree model.
  
  \item \textbf{GPU hardware and memory management} - Access to a high-memory GPU and the ability to manage its memory enabled us to push towards a larger model on a denser physical grid.  We used a 40-GB NVIDIA A100 GPU, which has significantly more memory than the 16-GB (or 32-GB) Tesla V100 used in \cite{weyn2020_cnn_cubed_sphere} and the 11-GB GTX 2080 used in \cite{rasp2021_resnet}.  Additionally, we took advantage of the gradient "checkpointing" (aka "rematerialization") features of \texttt{jax.remat}\footnote{\url{https://jax.readthedocs.io/en/latest/jax.html##jax.checkpoint}} and \texttt{hk.remat}\footnote{\url{https://dm-haiku.readthedocs.io/en/latest/api.html##haiku.remat}} to reduce instantaneous GPU memory usage and thereby compute a loss that accumulates over many rollout steps.
  
  \item \textbf{Loss} - There is quite a lot of freedom in how one reduces a multi-dimensional array of MSE values (with dimensions of latitude, longitude, pressure level, physical variable, rollout time step) to a single, scalar loss $\mathcal{L}$, and thereby calculate its gradient, $\nabla_{\theta}\mathcal{L}$, to train the model.  When using a weighted sum over this MSE array, as we did in this work, the question becomes, what is the relative importance of a particular physical variable, latitude range, pressure level, etc. when it comes to the overall, long-term forecast performance?  In our experiments we found that a simple heuristic --- re-scaling the data by the standard deviation of its temporal difference, as described in Section~\ref{sec:loss} --- worked significantly better than, say, re-scaling the data by its standard deviation.  As an aside we note that, while it may be tempting to replace this heuristic with a a more end-to-end-learned approach, (i) you would still have to use human judgement to pick a metric to optimize (e.g. globe-averaged $Z500$ at a 10-day forecast horizon) and (ii) directly optimizing over tens of rollout steps might not be effective \citep{metz2021_gradients_are_not_all_you_need}, even if you are able to fit the gradient into GPU memory.
\end{itemize}

\section{Conclusion}
\label{sec:conclusion}

We have presented a data-driven, machine learning approach for forecasting global weather using graph neural networks.  The system uses local information to step forward the current 3D atmospheric state by six hours, and multiple steps are chained together to produce skillful forecasts going out several days into the future.  The model works well, with forecast performance improving upon previous data-driven approaches and comparable to operational, physical models like GFS and ECMWF when evaluated on 1-degree scales and when using reanalysis data for the initial conditions.  We have also demonstrated that this model can be connected to live, operational forecasts from GFS to produce a hybrid physics+ML system that anticipates the output of the physics-based GFS model.

We see several directions for future work including: (i) pushing to yet finer spatial resolution, e.g. 0.25 degrees (we have some concerns that our 1.0-degree training data contains aliased, high-frequency information that is essentially impossible for our model to predict); (ii) using adaptive mesh refinement to optimize forecast performance given a fixed compute budget; (iii) using this data-driven model to cheaply generate large ensembles \citep{weyn2021_subseasonal_ensemble}; (iv) creating a data-driven data assimilation model that maps from observations to a grid of variables similar to the grid used in this work; and (v) most ambitiously, creating an end-to-end, data-driven forecast system \citep{schultz2021_can_deep_learning_beat_nwp} that maps from observations to observations with a minimal amount of physical priors imposed.  We believe that such a system could, in principle, outperform existing physics-based forecast systems because it could learn to predict effects that are not fully or optimally encoded into the physical models.

We believe that this work demonstrates the power of data-driven techniques for modeling the dynamics of physical systems, and we hope that it can stimulate follow-up efforts that use machine learning in the service of weather forecasting.

\section*{Acknowledgements} We would like to thank Stephan Rasp for valuable feedback on this work and manuscript.

\bibliographystyle{plainnat}
\bibliography{bib}

\end{document}